\documentclass[reprint,amsmath,amssymb,aps,prb]{revtex4-1}

\pdfoutput=1

\usepackage{hyperref,url}
\usepackage{amsmath}
\usepackage{amsfonts}
\usepackage[capitalise]{cleveref}
\usepackage{placeins}
\hypersetup{breaklinks,colorlinks=true,linkcolor=blue,citecolor=blue,urlcolor=blue,filecolor=blue}
\usepackage{graphicx}
\usepackage{dcolumn}
\usepackage[version=3]{mhchem}
\usepackage[english]{babel}
\usepackage{filecontents}
\usepackage{makecell}
\usepackage{graphicx}
\usepackage{courier}
\usepackage{amssymb,amsmath,amsthm,mathrsfs,comment,afterpage,accents}
\usepackage{mathtools}
\usepackage{braket}
\usepackage{mathrsfs}
\usepackage{courier}
\usepackage{color}
\usepackage{subdepth}
\usepackage{tablefootnote}
\usepackage{multirow}
\usepackage{booktabs}
\usepackage{siunitx}

\newcolumntype{?}{!{\vrule width 1pt}}

\newcommand\mat\mathbf
\newcommand\tr{\operatorname{tr}}
\newcommand{\joonho}[1]{{\textcolor{black} {#1}}}

\begin{document}

\author{Joonho Lee}
\email{linusjoonho@gmail.com}
\affiliation{College of Chemistry, University of California, Berkeley, California 94720, USA.}
\author{Luke W. Bertels}
\email{lwbertels@gmail.com}
\affiliation{College of Chemistry, University of California, Berkeley, California 94720, USA.}
\author{Martin Head-Gordon}
\email{mhg@cchem.berkeley.edu}
\affiliation{College of Chemistry, University of California, Berkeley, California 94720, USA.}

\title{Kohn-Sham Density Functional Theory with Complex, Spin-Restricted Orbitals:
%A New Route to Escape the Symmetry Dilemma
Accessing a New Class of Densities
without 
the Symmetry Dilemma
}
\begin{abstract}
We show that using complex, spin-restricted orbitals (cR) in Kohn-Sham density functional theory (KS-DFT)
allows one to access a new class of densities
that is not accessible by either spin-restricted (RKS) or spin-unrestricted (UKS) orbitals.
We further show that the real part of a cRKS density matrix can be non-idempotent when the imaginary part of the density matrix is not zero. 
Using cRKS orbitals shows  significant improvements in the triplet-singlet gaps of the TS12 benchmark set for the SPW92, PBE, BLYP, and TPSS functionals.
Moreover, it was shown that RKS and UKS yield qualitatively wrong charge densities and spin densities, respectively, leading to worse energetics.
We demonstrate that the modern functionals SCAN, MN15-L, and B97M-V show surprisingly no improvement even with a qualitatively more accurate density from cRKS orbitals. To this end, our work not only paves the way to escape
the symmetry dilemma whenever there exists a cRKS solution, but also suggests a new route to design better approximate XC functionals.
\end{abstract}
\maketitle
%<<<<<<< HEAD
%In this Letter, we show that 
%for certain systems the use of real, spin-restricted orbitals is too restrictive when searching for a density matrix, $\mathbf P$, as a solution to the Eurler-Lagrange equation of Kohn-Sham density functional theory (KS-DFT) even with the exact exchange-correlation (XC) functional. Furthermore, we suggest a new way to access density that is not accessible by more widely used KS-DFT schemes such as spin-restricted KS-DFT (RKS) and spin-unrestricted KS-DFT (UKS).
%In Kohn-Sham density functional theory (KS-DFT), spin-unrestricted KS-DFT (UKS) behaves better than spin-restricted KS-DFT (RKS) for systems with antiferromagnetically-coupled open-shell electrons at the cost of spin-polarization of the wavefunction.\cite{Gunnarsson1976,Seminario1994,Davidson1998}
{\it Introduction} In Kohn-Sham density functional theory (KS-DFT), it is clear that for systems with antiferomagnetically coupled open-shell electrons, spin-unrestricted KS-DFT (UKS) behaves better than spin-restricted KS-DFT (RKS) at the expense of spin-polarization.\cite{Gunnarsson1976,Seminario1994,Davidson1998}
%For finite-sized systems, this spin-polarization arises from the lack of strong (or static) correlation in a single-determinant wavefunction.\cite{Lykos1963,Perdew1995} 
For finite-sized systems, this spin-polarization is a manifestation of the lack of strong (or static) correlation in a single-determinant wavefunction.\cite{Lykos1963,Perdew1995}
This is known as a symmetry dilemma in KS-DFT, so called because exact solutions for finite systems should not break spin-symmetry yet the unphysical, spin-symmetry-broken approximate solutions often provide lower energies than symmetric approximate solutions.
%This is known as a symmetry dilemma in KS-DFT. In essence, it is a dilemma because exact solutions should not break spin-symmetry. Therefore, any spin-symmetry breaking should be considered {\it unphysical} for finite-sized systems.

As an attempt to understand this symmetry dilemma, Perdew and co-workers proposed the use of ``on-top'' pair density,\cite{Perdew1995,Perdew1997} $\Pi(\mathbf r)$, which is defined as
\begin{equation}
\Pi(\mathbf r) = \rho_\alpha (\mathbf r) \rho_\beta (\mathbf r).
\label{eq:pi}
\end{equation}
Instead of working with spin density variables, $\rho_\alpha(\mathbf r)$ and $\rho_\beta(\mathbf r)$, 
the fundamental variables are now the charge density, $\rho(\mathbf r) = \rho_\alpha(\mathbf r)+\rho_\beta(\mathbf r)$, 
and the magnetization density, $m(\mathbf r)$, which is a function of $\rho(\mathbf r)$ and $\Pi(\mathbf r)$.
One can accomplish the one-to-one mapping between $m(\mathbf r)$ and $\Pi(\mathbf r)$ following
\begin{equation}
m(\mathbf r) = \rho(\mathbf r) \left(1 - \frac{4\Pi(\mathbf r)}{\rho(\mathbf r)^2}\right)^{1/2}
\label{eq:m}
\end{equation}
Evidently, inserting Eq. \eqref{eq:pi} into Eq. \eqref{eq:m} yields a more familiar definition of the magnetization density,
$m(\mathbf r) = \rho_\alpha(\mathbf r) - \rho_\beta(\mathbf r)$. 
With $\rho(\mathbf r)$ and $m(\mathbf r)$, one can back out $\rho_\alpha(\mathbf r)$ and $\rho_\beta(\mathbf r)$ and these can be used to compute the
exchange-correlation (XC) potential, $v_\text{xc}(\mathbf r)$.
This framework allows one to treat $m(\mathbf r)$ as an auxiliary variable that is closely related to the on-top pair density. Therefore, it helps to explain
%<<<<<<< HEAD
spin-polarization in KS-DFT in terms of the on-top pair density of a wavefuntion. 
While this is certainly useful in understanding this symmetry dilemma, it is somewhat unsatisfying that
%=======
%spin-polarization in KS-DFT in terms of the wavefuntion language. While this is certainly useful in interpreting symmetry dilemma, it is somewhat unsatisfying that
%>>>>>>> 09e63a0e34b7bdacd31391cbf0a6aa914645b9d7
one has to interpret the magnetization density merely as an auxiliary variable.
%This approach is useful in understanding the symmetry dilemma, 
Moreover, it does not remedy
problems arising from evaluating $v_\text{xc}(\mathbf r)$ with spin densities. 
%After all, an $S=0$ state should have $m(\mathbf r) = 0$ for any $\mathbf r \in R^3$.

%<<<<<<< HEAD
In this Letter, 
we describe a way to access a new class of charge and spin densities that is not accessible by spin-polarization.
This is achieved by breaking time-reversal symmetry and complex symmetry of the KS-DFT determinant.\cite{Fukutome1981,Stuber2003,Small2015,Lee2019}
%=======
%Our goal is to illustrate another route to escape a particular symmetry dilemma without introducing any auxiliary variables or any symmetry breaking in $\rho$, $\rho_\alpha$, or $\rho_\beta$.
%We will focus on describing $S=0$ states where the use of RKS yields qualitatively wrong $\rho$.
%Even with the exact XC functional, the RKS determinant has to break the underlying spatial symmetry and yields qualitatively incorrect $\rho$.
%Instead, one can use complex RKS (cRKS) determinants to obtain qualitatively correct $\rho$ and also improve the
%triplet-singlet gap ($\Delta E_{S-T}$) significantly.
%In cRKS, the KS-DFT determinant breaks complex and time-reversal symmetry by allowing orbitals to complexify while preserving spin-symmetry.\cite{Fukutome1981,Stuber2003,Small2015,Lee2019}
%>>>>>>> 09e63a0e34b7bdacd31391cbf0a6aa914645b9d7
We will refer to this symmetry breaking as ``complex-polarization''.
For semi-local functionals, we show that this complex-polarization does not pose any dilemma in obtaining charge and spin densities.
%we illustrate another route to escape a particular symmetry dilemma without introducing any auxiliary variables or any symmetry breaking in $\rho$, $\rho_\alpha$, and $\rho_\beta$.
%We will focus on describing $S=0$ states where the use of RKS yields qualitatively wrong $\rho$.
%Even with the exact XC functional, the RKS determinant has to break the underlying spatial symmetry and yields qualitatively incorrect $\rho$.
%Instead, one can use complex RKS (cRKS) determinants to obtain qualitatively correct $\rho$ and also improves the
%triplet-singlet gap ($\Delta E_{S-T}$) significantly.
%In cRKS, the KS-DFT determinant breaks complex and time-reversal symmetry by allowing orbitals to complexify while preserving spin-symmetry.\cite{Fukutome1981,Stuber2003,Small2015,Lee2019}
%As we will see, complex-polarization in the cRKS determinant does not mean symmetry breaking in $\rho$.
We illustrate how this is achieved
and demonstrate
the numerical performance of popular semi-local functionals on chemical systems where complex-polarization
is {\it essential} to obtain correct charge and spin densities.
%Therefore, this provides a new way to escape the symmetry dilemma in KS-DFT.
%In particular, we break symmetry only in the KS-DFT determinant and do not break any symmetry in the real-space density.

{\it Theory} We start from the KS-DFT Lagrangian,\cite{Taube2010}
\begin{align}\nonumber
\mathcal L[\mathbf P] =& E_\text{KS-DFT}[\mathbf P] + \mu\left(\tr(\mathbf P\right) - \frac N2) + \tr(\mathbf A(\mathbf P^2 - \mathbf P))\\
&+ \tr (\mathbf B (\mathbf P^\dagger - \mathbf P))
\label{eq:lag}
\end{align}
where $N$ is the number of electrons, $\mathbf P$ is a one-particle density matrix (1PDM), $\mu$, $\mathbf A$, and $\mathbf B$ are Lagrange multipliers for constraining the trace, idempotency, and hermiticity of $\mathbf{P}$, respectively.
%of $\mathbf P$ and enforcing the  idempotency and hermiticity of $\mathbf P$. 
The idempotency of $\mathbf P$ guarantees the non-interacting nature of the KS-DFT problem.
For simplicity, we also assume that the computational basis is orthogonal.

These three constraints are naturally imposed by the definition of $\mathbf P$,
\begin{equation}
\mathbf P = \mathbf{C}_\text{occ} \mathbf{C}_\text{occ}^\dagger
\end{equation}
where $\mathbf{C}_\text{occ}$ is the occupied molecular orbital (MO) coefficient matrix.
However, when minimizing Eq. \eqref{eq:lag} with respect to $\mathbf P$ directly as is done in linear-scaling DFT approaches,
it is necessary to consider these constraints explicitly or impose them through density matrix purification techniques.\cite{Kussmann2013}
We write $\mathbf P$ in terms of its real and imaginary components,
\begin{equation}
\mathbf P = \mathbf X + i \mathbf Y
\end{equation}
We can then rewrite these constraints as follows:
the trace condition leads to
\begin{align}
\label{eq:trX}
\tr(\mathbf X) &= \frac N2\\
\label{eq:trY}
\tr(\mathbf Y) &= 0,
\end{align}
the idempotency becomes
\begin{align}
\label{eq:idemX}
\mathbf X &= \mathbf X^2 - \mathbf Y^2\\
\label{eq:idemY}
\mathbf Y &= \mathbf X\mathbf Y + \mathbf Y \mathbf X,
\end{align}
and the hermiticity yields
\begin{align}
\label{eq:symX}
\mathbf X &= \mathbf X ^T \\
\label{eq:symY}
\mathbf Y &= -\mathbf Y ^T.
\end{align}
Based on Eq. \eqref{eq:symY}, we conclude that $\mathbf Y$ is antisymmetric and therefore it automatically satisfies Eq. \eqref{eq:trY}.
This is an important observation since $\rho(\mathbf r)$ is computed from $\mathbf P$ following
\begin{equation}
\rho(\mathbf r) = 
2\sum_{\mu\nu}
\eta_\mu(\mathbf r)
\eta_\nu(\mathbf r)
P_{\mu\nu}
\end{equation}
where $\eta_\mu(\mathbf r)$ is a real-valued computational basis function.
This real-valuedness is sufficient for studying bound states of molecules and indeed most basis sets for finite-sized systems are real-valued. 
%The analysis here can be generalized to complex-valued basis sets such as planewaves. 
Since $\eta_\mu(\mathbf r)\eta_\nu(\mathbf r)$ is a symmetric tensor, an antisymmetric tensor, $\mathbf Y$, does not contribute to $\rho(\mathbf r)$.
In other words, it is equivalent to write
$
\rho(\mathbf r) = 2\sum_{\mu\nu}
\eta_\mu(\mathbf r)
\eta_\nu(\mathbf r)
X_{\mu\nu}
$.

The key insight here is that
by having non-zero $\mathbf Y$,
the idempotency constraint on $\mathbf X$ can be relaxed to 
Eq. \eqref{eq:idemX} and Eq. \eqref{eq:idemY}.
In other words, $\mathbf {X}$ does not have to satisfy 
$\mathbf X = \mathbf X^2$ as long as $\mathbf Y$ is non-negligible.
Such a density matrix $\mathbf P$ with $\tr(\mathbf Y^T \mathbf Y)\ne0$
is referred to as a
``fundamentally complex'' density matrix.
It is fundamentally complex in the sense that no unitary rotation of a real-valued MO coefficient matrix
in the complex plane 
can represent a fundamentally complex density matrix.\cite{Small2015}
Consequently, the use of fundamentally complex density matrices results in a broader class of densities which 
may come from a non-idempotent density matrix $\mathbf X$.
It is not obvious whether lifting this idempotency constraint on $\mathbf X$ would always yield a lower energy 
solution.
In fact, the energy lowering turns out to be quite rare to observe and systems with a high point group symmetry tend to
exhibit this energy lowering.

For semi-local functionals,  $E_\text{KS-DFT}$ depends only on $\mathbf X$ and reads
\begin{align}\nonumber
%E_\text{KS-DFT}[\mathbf P]
%=
E_\text{KS-DFT}[\mathbf X]
=&\:\:
E_T\left[\mathbf X\right]
+ E_V\left[\rho(\mathbf r)\right]
+ E_J\left[\rho(\mathbf r)\right]\\
&+ E_{XC} \left[\rho(\mathbf r), \nabla\rho(\mathbf r), \cdot\cdot\cdot\right]
+ E_{nn}
\label {eq:eks}
\end{align}
where
the kinetic energy is defined as
\begin{equation}
E_T\left[\mathbf X\right] = 
2\tr \left(
\mathbf X \mathbf T
\right)
\end{equation}
with $\mathbf T$ is the kinetic energy matrix in the computational basis ($T_{\mu\nu} = \langle\mu|-\frac12\nabla^2|\nu\rangle$),
the nuclear-electron attraction energy is
\begin{equation}
E_V\left[\rho(\mathbf r)\right]
=
-\sum_{I} Z_I\int_\mathbf{r}\frac{\rho(\mathbf r)}{||\mathbf r - \mathbf R_I||_2}
\end{equation}
with $I$ denoting the nuclei (or ions), the electron-electron repulsion energy reads
\begin{equation}
E_J\left[\rho(\mathbf r)\right]
=
\frac12\int_{\mathbf{r}_1}\int_{\mathbf{r}_2}\frac{\rho(\mathbf r_1)\rho(\mathbf r_2)}{||\mathbf r_1 - \mathbf r_2||_2},
\end{equation}
$E_{XC}$ denotes the XC energy, and $E_{nn}$ is the nuclear-nuclear repulsion energy.
We emphasize that as far as Eq. \eqref{eq:eks} is concerned
there is no symmetry breaking associated with $\mathbf X$ and $\rho$.
Furthermore, there is no auxiliary variable introduced in the energy evaluation.

\joonho{Beyond permitting access to densities not describable by RKS and even UKS, one other point should be mentioned.
Just as $\mathbf X$ is non-idempotent, the cRKS determinant is inherently multiconfigurational (MC) as shown in ref. \citenum{Small2015}.
A single complex orbital $\xi$ is parametrized by $\theta$ to have an arbitrary linear combination of real $\eta$ and imaginary $\bar{\eta}$ orbitals, 
%\begin{equation}
$
\xi = \cos(\theta) \eta - i \sin(\theta) \bar{\eta}.
$
%\end{equation}
A two-electron closed-shell determinant made of a complex spatial orbital $\xi_i$ follows
\begin{equation}
|\Psi \rangle = \mathcal A [\xi\xi\left(\alpha\beta\right)] = \mathcal A [\left(\Pi + \Omega\right)\left(\alpha\beta\right)]
\label{eq:wfn}
\end{equation}
where $\mathcal A$ is the antisymmetrizer including a normalization factor, $\Pi = \cos^2(\theta)\eta\eta - \sin^2(\theta)\bar{\eta}\bar{\eta}$, and
$\Omega = -i\sin(\theta)\cos(\theta)(\eta\bar{\eta}+\bar{\eta}\eta)$.
These two spatial wavefunctions, $\Pi$ and $\Omega$, highlight the MR character of a cRKS single-determinant $|\Psi\rangle$.
$\Pi$ is a two-configuration wavefunction (lower energy) and $\Omega$ is a open-shell wavefunction (higher energy).
The competition between energy-lowering via $\Pi$ and an energy penalty through $\Omega$ determines whether an RKS electron pair complexifies to cRKS.
In the $\eta$ basis, the corresponding 2x2 matrix, $\mathbf X$ is
\begin{equation}
\mathbf X  = 
\begin{bmatrix}
\cos^2(\theta) & 0 \\
0 &  \sin^2(\theta)
\end{bmatrix}
\end{equation}
$\theta$ = 0 represents an RKS density matrix whereas $\theta = \pi/4$ yields two half-occupied orbitals. Depending on $\theta$, it is possible to obtain a non-idempotent $\mathbf X$.
While the hidden MC form in the cRKS determinant is limited, its real 1PDM (for $E_T$) and its $\rho(\mathbf r)$ (for $E_J$ and $E_V$) can all be consistently
evaluated using existing XC functionals. Thus cRKS can be viewed as a limited case of MC-DFT without the formal challenges\cite{Miehlich1997,grafenstein1998density,Kurzweil2009}
and practical double-counting problems\cite{Miehlich1997,Gusarov2004,Khait2004,LiManni2014} normally associated with that field.}

{\it Results} We evaluate the numerical performance of RKS, UKS, and cRKS on the recently developed TS12 benchmark set.\cite{Lee2019}
%Numerically, cRKS indeed provides quantitatively more accurate
%results
%compared to RKS and UKS when used with some semi-local XC functionals.
This data set contains the experimental singlet-triplet gaps of 12 atoms and molecules:
%The particular benchmark set we chose is called TS12 and involves a total of
%12 data points for the triplet-singlet gaps of atoms and diatomic molecules.\cite{Lee2019}
%The TS12 set includes 
C, NF, NH, NO$^-$, O$_2$, O, PF, PH, S$_2$, S, Si, and SO.
The ground states of these molecules are triplets.
The lowest singlet states for each system are singlet biradicals and exhibit a spin-restricted Hartree-Fock (RHF) to complex RHF (cRHF) instability.
%Every singlet state in the TS12 set exhibits the
%spin-restricted Hartree-Fock (RHF) to complex RHF (cRHF) instability.
This instability is driven by the underlying point group symmetry which gives rise to the degeneracy between the highest occupied MO (HOMO) and the lowest unoccupied MO (LUMO).

The correct lowest singlet states of these systems should singly-occupy both HOMO and LUMO and thereby obtain charge and spin densities that obey the underlying point group symmetry.
Even with the exact XC functional,
RKS orbitals are qualitatively wrong in these cases as they doubly-occupy the HOMO, breaking the point group symmetry of the system and 
leading to a broken-symmetry charge density. 
%RKS orbitals are qualitatively wrong in these cases because they doubly-occupy the HOMO which breaks the point group symmetry.
%This spatial symmetry breaking further leads to a broken-symmetry charge density. 
Since the exact XC functional yields not only exact energy but also exact charge density, 
accessing a different class of density matrix other than those from RKS orbitals is necessary.
A similar phenomenon was first pointed out by Pople, Gill, and Handy in the context of spin-polarization in open-shell systems.\cite{pople1995spin}
Indeed, UKS can achieve these single occupations by breaking spin-symmetry.
However, we will show that spin-polarization is not the way to access the right charge and spin densities in the systems in TS12.
%as evidenced by the triplet-singlet gaps.

%UKS can achieve these single occupations at the expense of the spin-polarization symmetry dilemma.
%The correct solution should singly-occupy both HOMO and LUMO and thereby obtains the correct point group symmetry of the lowest singlet state without breaking spin-symmetry. 
%It is expected that a well-designed approximate XC functional should exhibit this instability
%with the single occupations in the HOMO and LUMO.

We investigate a local density approximation (LDA) functional, SPW92,\cite{Hohenberg1964,Kohn1965,Slater1963,Perdew1992} two generalized gradient approximation (GGA) functionals, BLYP\cite{Becke1988,Lee1988} and PBE,\cite{Perdew1996} and four meta-GGA (mGGA) functionals, TPSS,\cite{Tao2003} SCAN,\cite{Sun2015} 
MN15-L,\cite{haoyu2016mn15} and B97M-V.\cite{mardirossian2015mapping} 
We used aug-cc-pVQZ basis set,\cite{Dunning1989,Kendall1992} 99 points for XC radial quadrature and 590 points for XC angular quadrature. We only report KS-DFT solutions found to be locally stable based on stability analysis.\cite{Bauernschmitt1996,Sharada2015} The triplet ground states are computed with UKS ($M_S=1$) and we focus on the symmetry dilemma of the singlet states. All calculations were performed with a development version of Q-Chem.\cite{Shao2015}

\begin{table}[h!]
  \centering
  \begin{tabular}{|c|r|r|r|r|}\hline
 & \multicolumn{1}{c|}{SPW92} & \multicolumn{1}{c|}{PBE} & \multicolumn{1}{c|}{BLYP} & \multicolumn{1}{c|}{TPSS} \\ \Xhline{3\arrayrulewidth}
& \multicolumn{4}{c|}{RKS} \\ \hline
MSD & 10.85 & 13.09 & 9.94 & 13.85 \\ \hline
RMSD & 11.46 & 13.53 & 10.34 & 14.42 \\ \hline
%MAX & 19.17 & 21.19 & 16.58 & 23.43 \\ \hline
%MIN & 6.92 & 9.01 & 6.58 & 9.64 \\ \Xhline{3\arrayrulewidth}
& \multicolumn{4}{c|}{UKS} \\ \hline
% & USPW92 & UPBE & UBLYP & UTPSS \\ \Xhline{3\arrayrulewidth}
MSD & -13.70 & -16.57 & -17.61 & -17.26 \\\hline
RMSD & 14.46 & 17.61 & 18.64 & 18.19 \\ \hline
%MAX & 23.84 & 29.17 & 30.49 & 28.09 \\ \hline
%MIN & 7.95 & 8.79 & 9.31 & 9.28 \\  \Xhline{3\arrayrulewidth}
% & cRSPW92 & cRPBE & cRBLYP & cRTPSS \\ \Xhline{3\arrayrulewidth}
& \multicolumn{4}{c|}{cRKS} \\ \hline
MSD & -1.23 & 3.00 & 0.90 & 7.94 \\ \hline 
RMSD & 2.19 & 3.41 & 1.91 & 8.63\\\hline
%MAX & -4.42 & 5.27 & 3.76 & 14.30 \\ \hline
%MIN & -0.25 & -0.18 & -0.12 & 3.77 \\ \Xhline{3\arrayrulewidth}
  \end{tabular}
  \caption{
The deviation (kcal/mol) with respect to experimental values in $\Delta E_\text{T-S} (= E_S - E_T)$ of the SPW92, PBE, BLYP, and TPSS functionals obtained with restricted (RKS), unrestricted (UKS), and complex, restricted (cRKS) orbitals. 
RMSD stands for root-mean-square-deviation and MSD stands for mean-signed-deviation. 
%MAX and MIN are maximum and minimum deviations, respectively.
}
  \label{tab:st1}%\footnote{test}
\end{table}

In Table \ref{tab:st1}, we present the root-mean-square-deviation (RMSD) and mean-signed deviation (MSD) of the relatively old XC functionals SPW92, BLYP, PBE, and TPSS. Using RKS orbitals clearly overestimates the triplet-singlet gap as restricted orbitals cannot describe the open-shell nature of the lowest singlet state of these systems. These RKS energies for singlet states are too high so that the gap is overestimated, as evidenced by the positive MSDs. 
%The positive MSDs are attributed to this.
Using UKS orbitals has the opposite problem. Namely, the gaps are all underestimated. This is due to the undesired mixing of the singlet and triplet states (i.e., spin-contamination). Since the triplet state is lower in energy, the resulting UKS $M_S=0$ state energy is too low. The negative MSDs are due to this spin-contamination.
%The use of cRKS orbitals improves more than a factor of 5 in terms of both MSD and RMSD in the case of SPW92, PBE and BLYP. 
The use of cRKS orbitals improves both the MSD and RMSD by more than a factor of 5 in the case of SPW92, PBE and BLYP.
The improvement in TPSS is less impressive although it is still a factor of 2 improvement.

This significant improvement is simply due to obtaining $\rho$ from a non-idempotent $\mathbf X$ {\it without spin-polarization}. 
Indeed, in all systems in TS12, the resulting $\mathbf X$ has exactly two eigenvalues of 0.5 when obtained from these four XC functionals. These two eigenvalues represent the single-occupancy of
HOMO and LUMO which is qualitatively correct for these systems.
\begin{table}[h!]
  \centering
  \begin{tabular}{|c?r?r|r|r?r|r|r|}\hline
& $\Delta E_T$ & $\Delta E_V$ & $\Delta E_J$ & $\Delta E_\text{Coul}$ & $\Delta E_X$ & $\Delta E_C$ & $\Delta E_\text{XC}$\\ \Xhline{3\arrayrulewidth}
SPW92 & 12.89 & -45.70 & 11.65 & -34.05 & 7.88 & 0.20 & 8.09\\\hline
PBE & 10.69 & -39.57 & 7.06 & -32.51 & 12.13 & -1.38 & 10.76\\\hline
BLYP & 9.96 & -36.43 & 4.09 & -32.33 & 13.05 & -0.38 & 12.67\\\hline
TPSS & 5.95 & -26.95 & -2.28 & -29.22 & 18.45 & -1.65 & 16.80\\\hline
  \end{tabular}
  \caption{
The energy differences (kcal/mol) between cRKS and RKS solutions of \ce{O2} (i.e., $\Delta E = \Delta E_\text{cRKS} - \Delta E_\text{RKS}$) 
for each component in Eq. \eqref{eq:eks}.
$\Delta E_\text{Coul} = \Delta E_J + \Delta E_V$,  $\Delta E_X$, and $\Delta E_C$ are the energy differences in classical Coulomb energy, exchange energy, 
and correlation energy, respectively. %The data for other functionals and molecules are available in Supplemental Material.
}
  \label{tab:st2}%\footnote{test}
\end{table}
To understand what drives the energy lowering from RKS to cRKS, we decompose the energy difference between RKS and cRKS into individual
energy contributions. We present this energy decomposition for \ce{O2} in Table \ref{tab:st2}.
The qualitative behavior discussed here holds for other systems as well.
All four XC functionals show an increase in the kinetic energy as well as the XC energy.
The increase in the XC energy is driven by the increase in the exchange energy. The correlation energy shows only a small change.
The RKS to cRKS instability is driven by the classical Coulomb energy.
In particular, a significant energy lowering in the electron-nuclear attraction energy is the driving force of this instability in all systems in the TS12 set.

\begin{table}[h!]
  \centering
  \begin{tabular}{|c|r|r|r|}\hline
 & \multicolumn{1}{c|}{SCAN} & \multicolumn{1}{c|}{MN15-L} & \multicolumn{1}{c|}{B97M-V} \\ \Xhline{3\arrayrulewidth}
& \multicolumn{3}{c|}{RKS} \\ \hline
MSD & 19.50 & 10.64 & 11.82 \\ \hline
RMSD & 19.82 & 11.16 & 12.22 \\ \hline
%MAX & 26.97 & 16.79 & 18.73 \\ \hline
%MIN & 14.17 & 5.71 & 6.50 \\ \Xhline{3\arrayrulewidth}
& \multicolumn{3}{c|}{UKS} \\ \hline
MSD & -16.24 & -10.79 & -13.69 \\ \hline
RMSD & 17.74 & 11.97 & 14.71 \\ \hline
%MAX & 31.95 & 20.75 & 25.79 \\ \hline
%MIN & 7.04 & 4.28 & 7.25 \\ \Xhline{3\arrayrulewidth}
& \multicolumn{3}{c|}{cRKS} \\ \hline
MSD & 15.55 & 10.50 & 10.34 \\ \hline
RMSD & 16.39 & 11.00 & 11.15 \\ \hline
%MAX & 25.99 & 16.79 & 18.73 \\ \hline
%MIN & 7.88 & 5.71 & 4.55 \\ \Xhline{3\arrayrulewidth}
  \end{tabular}
  \caption{
Same as Table \ref{tab:st1} except we have the SCAN, MN15-L, and B97M-V functionals presented here.
Note that MN15-L and B97M-V have no stable cRKS solutions for a half of the data points (i.e., \ce{C}, \ce{NF}, \ce{NH}, \ce{NO-}, \ce{O2}, and \ce{O}).
}
  \label{tab:st3}%\footnote{test}
\end{table}

The results for the more modern mGGA functionals SCAN, MN15-L, and B97M-V, %show an alarming behavior.
are presented in Table \ref{tab:st3}.
%This is evident from Table \ref{tab:st3} where the use of cRKS orbitals does not improve the triplet-singlet gaps.
The RKS and UKS results are qualitatively similar to what was observed in the above four functionals.
Namely, the overestimation in gaps was observed with RKS while the underestimation in gaps was observed with UKS.
SCAN with RKS orbitals shows a particularly poor behavior compared to all the other functionals examined in this work.
SCAN becomes comparable to other functionals when used with UKS orbitals.
The cRKS results with these three modern functionals are rather surprising due to a qualitatively different behavior from that of the four functionals discussed before.
The use of cRKS orbitals for these functionals does not improve the quantitative energetics.
Furthermore, MN15-L and B97M-V 
exhibit no stable cRKS solutions for \ce{C}, \ce{NF}, \ce{NH}, \ce{NO-}, \ce{O2}, and \ce{O}.
Additionally, MN15-L shows occupation numbers of (0.61, 0.39) and (0.78, 0.22) for \ce{PF} and \ce{PH}, respectively.
%In addition to this, MN15-L shows (0.61, 0.39) and (0.78, 0.22) occupations numbers for \ce{PF} and \ce{PH}, respectively.
Non-idempotent $\mathbf X$ is essential to describe the open-shell nature of singlet ground states in these systems.
Therefore, with this improved density matrix and $\rho$ therefrom, approximate XC functionals should perform better.
The poor performance of these modern mGGA functionals with cRKS orbitals suggests that
\joonho{data such as the TS12 set may be useful for XC functional development.}
%In particular, well-behaved functionals should improve the quantitative accuracy for describing singlet states in TS12 when combined with cRKS orbitals.

Lastly, we present density plots to compare the qualitative differences between densities from RKS and cRKS orbitals.
As mentioned throughout this Letter, 
RKS orbitals break the point group spatial symmetry of systems in the TS12 set whereas cRKS orbitals preserve the symmetry.\cite{footnote1}
The symmetry breaking at the orbital level results in the symmetry breaking in $\rho$.
We shall illustrate this point by looking at $\rho$ represented on a real-space grid.

\begin{figure}[h!]
\includegraphics[scale=0.4]{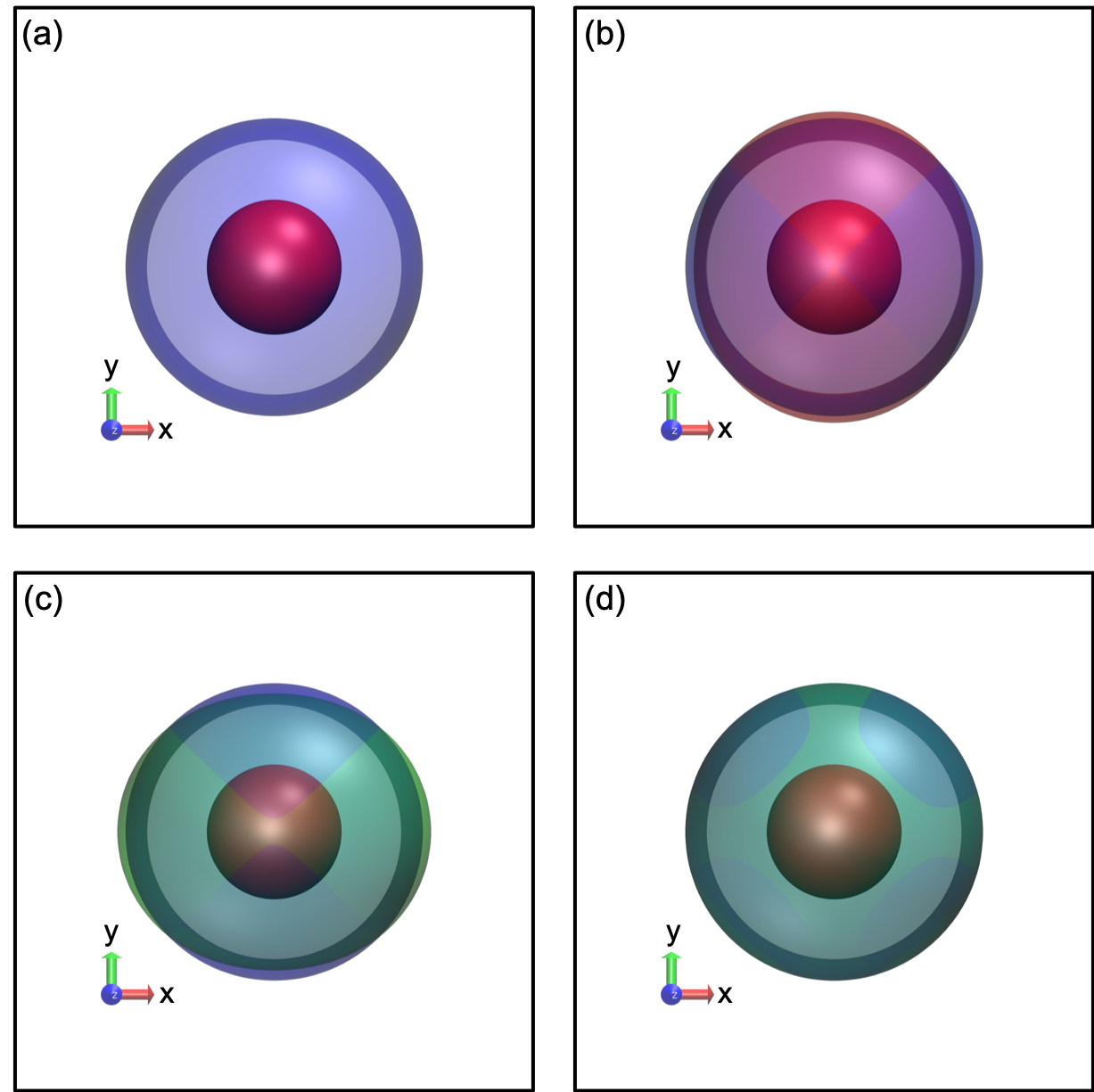}
\caption{
Density ($\rho(\mathbf r)$) represented on a real-space grid
for \ce{O2} computed with the BLYP XC functional.
\ce{O2} is aligned along the bond axis (z-axis)
so that we can inspect the cylindrical symmetry of density.
(a) cRKS density (blue),
(b) RKS (red) density superimposed on cRKS density (blue),
(c) UKS (green) two times $\alpha$-spin density ($2\rho_\alpha(\mathbf r)$) superimposed on cRKS density (blue),
and
(d) UKS (green) charge-density superimposed on cRKS density (blue).
The factor of 2 in (c) is to account for the difference between normalization factors of charge and spin densities.
Every plot is based on an isosurface value of 0.08 au.
The $z$-axis is pointing out of the page.
}
\label{fig:o2}
\end{figure}

In Figure \ref{fig:o2}, 
we present the real-space density of \ce{O2} computed with BLYP. 
The discussion also applies to other systems in the TS12 set and real-space densities are qualitatively similar across different XC functionals.
\ce{O2} is cylindrically symmetric around its interatomic axis.
Therefore, qualitatively correct density should exhibit this symmetry. 
This is found to be true with the cRKS density in Figure \ref{fig:o2} (a). 
However, in Figure \ref{fig:o2} (b), the RKS density is elongated along y-axis
and overall not cylindrically symmetric.
This is due to the spatial symmetry breaking caused by doubly occupying HOMO which mixes two low-lying singlet states, $^1\Delta_g$ and $^1\Sigma_g^+$.\cite{Small2015}
The UKS $\alpha$-spin density also breaks the cylindrical symmetry to a smaller extent than does RKS and is elongated along $x$-axis.
%It is interesting that t
The UKS charge density (i.e., the sum of $\alpha$ and $\beta$ density) is only $x$-$y$ symmetric and breaks the cylindrical symmetry.
The $x$-$y$ symmetry is because the UKS $\beta$-spin density is rotated 90$^\circ$ from the $\alpha$ density (i.e., elongated along $y$ axis).

The qualitatively improved UKS charge density (compared to the RKS one) is attractive though it is still qualitatively wrong.
The on-top pair density interpretation provides a good way to understand
the spin-symmetry breaking in this case, but there is no quantitative benefit.
%One may hope that since the charge-density is qualitatively correct the KS-DFT energy is also quantitatively accurate.
The resulting energetics from the UKS spin-densities are far from chemical accuracy (1 kcal/mol) as it is clear in Tables \ref{tab:st1} and \ref{tab:st2}.
In particular, due to the spatial symmetry breaking in each spin density and charge density, the KS-DFT XC energy evaluated with these spin densities
are inadequate to reliably estimate the triplet-singlet gaps. In contrast to this, we observe a qualitatively correct charge density (and also a spin density which is a half of the charge density) with the cRKS determinant. At the same time, the resulting cRKS triplet-singlet gaps are improved as it is shown in Table \ref{tab:st1}.

{\it Conclusions} In this Letter, we showed that it is possible to access a different class of (both charge and spin) densities that is not possible to obtain
within either RKS or UKS.
This is achieved by using cRKS to obtain the density. Although the cRKS determinant breaks time-reversal symmetry and complex symmetry, the resulting densities do not exhibit this symmetry breaking. Furthermore, based on the triplet-singlet test set (TS12), we showed that the cRKS charge densities follow the point group symmetry while RKS does not. This allows cRKS to improve the quantitative accuracy of some XC functionals by a factor of 5. 
We also showed that the UKS charge density is qualitatively incorrect and the resulting energies are far from chemical accuracy due to spin-contamination.
Lastly, we note that modern mGGA functionals (SCAN, MN15-L, and B97M-V) do not show any significant improvements even when correct densities from cRKS are used.
Even with the exact XC functional, one needs a cRKS determinant to obtain qualitatively correct charge and spin densities for the systems considered here. 
This Letter suggests that these modern functionals might lack some aspects of the exact XC functional.
We hope that our study 
\joonho{provides a new class of data that can be used to assess, and possibly inform the design of new approximate XC functionals.}
\joonho{The key benefit is a route to escape
the symmetry dilemma whenever complex-polarization is relevant.}

{\it Acknowledgement} This research was supported by the Director, Office of Science, Office of Basic Energy Sciences, of the U.S. Department of Energy under Contract No. DE-AC02-05CH11231.
We thank Fionn Malone, Yuezhi Mao, and Narbe Mardirossian for helpful comments.
The raw data for TS12 (Tables \ref{tab:st1}, \ref{tab:st2}, \ref{tab:st3}) is available in the Supplemental Material.
\bibliography{main}

\begin{thebibliography}{36}
\bibitem [{foo()}]{footnote1}%
  \BibitemOpen
  \href@noop {} {\ }\bibinfo {note} {It turns out that for the atoms in TS12
  (i.e., C, O, S, and Si) some subtlety arises. The singlet ground state of
  these atoms is five-fold degenerate ($^{1}$D). The cRKS density is not
  spherically symmetric since it cannot describe this ground state degeneracy.
  However, we were able to obtain a non-symmetric density from exact
  wavefunction calculations (full configuration interaction) that is very
  similar to the cRKS density. If densities from these five degenerate
  solutions are averaged with equal weights, one obtains a spherically
  symmetric density. We conclude that the cRKS density is still qualitatively
  correct up to the ground state degeneracy.}\BibitemShut {Stop}%
\end{thebibliography}
\end{document}